\documentclass[reprint, amsmath, amssymb, aps]{revtex4-2}
\usepackage{tikz}
\usetikzlibrary{arrows.meta,decorations.pathmorphing}
\usetikzlibrary{decorations.pathmorphing,decorations.markings}
\usepackage{tikz}
\usetikzlibrary{decorations.pathmorphing, decorations.markings}
\usepackage{feynmp-auto}
\usepackage{graphicx}
\usepackage{dcolumn}
\usepackage{bm}

\begin{document}

\preprint{APS/123-QED}

\title{Joint Simulation Study of Chiral Symmetry Recovery and Transport Response in QCD at Finite Temperature and Chemical Potentials}

\author{Jingxu Wu}
\affiliation{Faculty of Physics, M.V. Lomonosov Moscow State University, Moscow 119991, Russia}

\author{Yuwei Yin}
\affiliation{ Institut Polytechnique de Paris, 91128 Palaiseau, France}

\author{Jianing Shi}
\affiliation{School of Energy Engineering, Xinjiang Institute of Engineering, Urumqi 830023, China}

\date{\today}

\begin{abstract}
We investigate QCD matter using the two-flavor PNJL model and find that the shear viscosity to entropy density ratio $\eta/s$ reaches a minimum while the bulk viscosity $\zeta/s$ peaks near the chiral crossover line, particularly at large baryon chemical potential, indicating sensitivity to critical dynamics and offering potential signatures for the QCD critical endpoint in heavy-ion collisions.
\end{abstract}

\maketitle


\section{Introduction}

Understanding the QCD phase transition and the nature of strongly interacting matter under extreme conditions remains a cornerstone of modern nuclear and particle physics. Relativistic heavy-ion collision experiments at RHIC and the LHC have provided compelling evidence for the formation of a deconfined quark-gluon plasma (QGP), characterized by collective flow, near-perfect fluidity, and thermalized hadron yields\cite{Adams:2005dq,Abelev:2012rv,Andronic:2018qqt,Gale:2013da,Romatschke:2007mq}. The theoretical description of such systems relies on lattice QCD simulations\cite{Aoki:2006we,Borsanyi:2010bp,Bhattacharya:2014ara,Bazavov:2014pvz,HotQCD:2019xop} as well as effective field theories including NJL-type models and Polyakov-loop extensions\cite{Fukushima:2003fw,Ratti:2005jh,Roessner:2006xn,Schaefer:2007pw}. These approaches have illuminated the crossover nature of the transition at vanishing baryon chemical potential and suggest the existence of a critical endpoint (CEP) at finite density\cite{Stephanov:2004wx,Stephanov:2009ra,Fukushima:2010bq}.

Despite major progress in the field, several crucial questions remain unresolved: What is the precise location of the QCD critical point? How do thermodynamic fluctuations influence transport properties near the phase boundary? Can signatures of the phase transition be unambiguously identified in experimentally observable quantities? These questions have stimulated a large body of theoretical and experimental work over the past two decades, particularly within the framework of the Beam Energy Scan (BES) program at RHIC\cite{Luo:2017faz}, the NA61/SHINE experiment at CERN\cite{NA61:2021dxf}, and upcoming facilities such as NICA and FAIR\cite{Fukushima:2020cmk,Bzdak:2019pkr}.

In parallel, lattice QCD has provided valuable insights into the thermodynamics of QCD matter at zero and small chemical potential. High-precision lattice simulations with physical quark masses have established that the QCD transition at $\mu_B = 0$ is a crossover, not a first-order phase transition\cite{Borsanyi:2010cj,Bazavov:2014pvz}. The equation of state (EoS), susceptibilities of conserved charges, and correlation functions of energy-momentum tensors have been computed with increasing accuracy\cite{HotQCD:2019xop,Bazavov:2020bjn,Astrakhantsev:2018oue}. However, due to the fermion sign problem, direct simulations at large baryon density remain computationally intractable, necessitating effective models for exploration of the full phase diagram.

The Polyakov–Nambu–Jona-Lasinio (PNJL) model has emerged as a powerful tool for studying QCD thermodynamics at finite $T$ and $\mu_B$\cite{Fukushima:2003fw,Ratti:2005jh,Sakai:2008py}. By combining chiral symmetry breaking via the NJL mechanism with confinement dynamics through the Polyakov loop, the PNJL model captures essential nonperturbative features of the QCD transition. Variants of the model have been used to map phase boundaries, study quark number susceptibilities, and analyze the behavior of order parameters near the critical region\cite{Costa:2010zw,Abuki:2008nm}. Importantly, the model can be extended to incorporate quark backreaction on the Polyakov potential, magnetic field effects, and vector interactions\cite{Ferreira:2014kpa,Gatto:2010qs,Bratovic:2012qs}.

Transport coefficients such as shear viscosity ($\eta$), bulk viscosity ($\zeta$), and thermal conductivity ($\kappa$) play a central role in characterizing the dynamical properties of the medium\cite{Jeon:1995zm,Arnold:2000dr}. These quantities control the response of the system to external perturbations and influence collective observables such as flow coefficients ($v_n$), HBT radii, and transverse momentum correlations\cite{Romatschke:2007mq,Schenke:2010rr,Gale:2013da}. Among these, $\eta/s$ has received the most attention due to its conjectured lower bound in strongly coupled quantum field theories\cite{Kovtun:2004de}, while $\zeta/s$ is of particular interest near the phase transition where conformal symmetry is strongly broken\cite{Karsch:2007jc,Monnai:2009ad,Kapusta:2012zb}.

From a theoretical standpoint, transport coefficients can be derived from both kinetic theory and diagrammatic approaches. In the relaxation time approximation (RTA), the shear and bulk viscosities are obtained from integrals over the distribution function weighted by energy-dependent collision times\cite{Dusling:2011fd,Marty:2013ita}. The results depend on the quasiparticle dispersion relation, the degeneracy factor, and the behavior of the entropy density. More refined techniques involve Kubo relations, where transport coefficients are extracted from the low-frequency limit of retarded Green's functions of conserved currents or energy-momentum tensors\cite{Aarts:2002cc,Ghiglieri:2018dib,Astrakhantsev:2018oue}. These approaches reveal the interplay between spectral density modifications and medium effects near the crossover.

In recent years, Bayesian analysis frameworks have been developed to extract temperature-dependent $\eta/s(T)$ and $\zeta/s(T)$ from experimental data by comparing event-by-event hydrodynamic simulations with measured observables\cite{Bernhard:2019bmu,JETSCAPE:2020mzn,Parkkila:2022rlm,Nijs:2020ors}. These analyses suggest a non-monotonic behavior of bulk viscosity with a peak near $T_c$, and a minimum of shear viscosity close to the phase boundary. Additionally, modern efforts have focused on the role of out-of-equilibrium fluctuations, critical slowing down, and non-Gaussian cumulants as probes of the CEP\cite{Stephanov:2009ra,Bluhm:2016byc,Bzdak:2019pkr,Nahrgang:2020yxm}. Dynamical models incorporating noise and stochastic terms in hydrodynamics have been formulated to capture these effects\cite{An:2021wof,Du:2020bxp}.

Experimental advances have further enhanced our ability to probe the QCD medium. The RHIC Beam Energy Scan Phase II (BES-II) program has improved statistics and detector capabilities for measuring higher-order cumulants, dilepton spectra, and directed flow at low energies\cite{STAR:2020tga,STAR:2021fge}. Similarly, the ALICE experiment has provided detailed measurements of anisotropic flow, strangeness enhancement, and correlations in Pb--Pb collisions at the LHC\cite{ALICE:2018vuu,ALICE:2020siw}. These results provide crucial constraints for theoretical models and highlight the need for transport predictions that are sensitive to both criticality and non-equilibrium dynamics.

In this work, we study the thermodynamic and transport properties of two-flavor QCD matter using the PNJL model with a focus on the temperature and baryon chemical potential dependence of $\eta/s$ and $\zeta/s$. We begin by outlining the theoretical framework, including the effective potential, gap equations, and the relaxation time formalism for transport coefficients. We then solve the model numerically on a dense grid in the $T$--$\mu_B$ plane to extract chiral order parameters, pressure, entropy density, and speed of sound. Using these quantities, we compute $\eta/s$ and $\zeta/s$, and explore their relation to the phase boundary. In addition, we provide diagrammatic representations for the relevant spectral functions and derive the transport integrals within the kinetic theory framework.

Our analysis reveals that $\eta/s$ develops a valley structure along the crossover line, while $\zeta/s$ exhibits a pronounced peak at low temperatures and moderate $\mu_B$, consistent with expectations near the CEP. These findings are supported by the behavior of the trace anomaly, speed of sound, and entropy gradients. We further discuss the implications for hydrodynamic modeling and identify potential observables that can signal the critical behavior in ongoing and future experiments. Through this study, we aim to provide a unified and comprehensive picture of the transport landscape of QCD matter under realistic conditions.

\section{Theoretical Framework}

This section presents a comprehensive derivation of the theoretical model employed to study both the thermodynamic and transport properties of strongly interacting QCD matter. The starting point is the two-flavor Polyakov-loop extended Nambu--Jona-Lasinio (PNJL) model, which incorporates both chiral symmetry dynamics and confinement-like effects. We derive the mean-field thermodynamic potential, the corresponding gap equations, and the Kubo formulas for transport coefficients under the relaxation time approximation (RTA). Finally, we formulate the microscopic link between the shear/bulk viscosity and the retarded correlators of the energy-momentum tensor.

\subsection{PNJL Lagrangian and Symmetry Structure}

The PNJL model is defined by the Euclidean Lagrangian:
\begin{equation}
\mathcal{L}_{\text{PNJL}} = \bar{q}(i\gamma^\mu D_\mu - m_0)q + G[(\bar{q}q)^2 + (\bar{q}i\gamma_5 \vec{\tau} q)^2] - \mathcal{U}(\Phi, \bar{\Phi}, T),
\end{equation}
where $q$ is the two-flavor quark field, $m_0$ the current quark mass, $G$ the scalar four-fermion coupling, and $\vec{\tau}$ the Pauli matrices. The Polyakov loop variables $\Phi$ and $\bar{\Phi}$ are defined via:
\begin{equation}
\Phi = \frac{1}{N_c} \mathrm{Tr}_c \left[ \mathcal{P} \exp\left(i\int_0^\beta A_4 d\tau\right) \right],
\end{equation}
and couple quarks to a temporal background gluon field, mimicking confinement.

The effective potential $\mathcal{U}(\Phi, \bar{\Phi}, T)$ is parametrized to reproduce pure-gauge lattice data, typically as:
\begin{equation}
\frac{\mathcal{U}}{T^4} = -\frac{a(T)}{2}\bar{\Phi}\Phi + b(T)\ln[1 - 6\bar{\Phi}\Phi + 4(\bar{\Phi}^3 + \Phi^3) - 3(\bar{\Phi}\Phi)^2].
\end{equation}

\subsection{Thermodynamic Potential and Gap Equations}

The grand potential in mean-field reads:
\begin{widetext}
\begin{align}
\Omega &= \mathcal{U}(\Phi, \bar{\Phi}, T) + \frac{(M - m_0)^2}{4G} - 2N_f N_c \int^{\Lambda} \frac{d^3p}{(2\pi)^3} E_p \\
&\quad - 2T N_f \int \frac{d^3p}{(2\pi)^3} \Big[ \ln(1 + 3\Phi e^{-(E_p - \mu)/T} + 3\bar{\Phi} e^{-2(E_p - \mu)/T} + e^{-3(E_p - \mu)/T}) \nonumber\\
&\qquad\qquad + \ln(1 + 3\bar{\Phi} e^{-(E_p + \mu)/T} + 3\Phi e^{-2(E_p + \mu)/T} + e^{-3(E_p + \mu)/T}) \Big],
\end{align}
\end{widetext}
with $E_p = \sqrt{p^2 + M^2}$ and constituent quark mass $M = m_0 - 2G\langle\bar{q}q\rangle$.

Gap equations follow from the stationarity conditions:
\begin{equation}
\frac{\partial \Omega}{\partial M} = 0, \quad \frac{\partial \Omega}{\partial \Phi} = 0, \quad \frac{\partial \Omega}{\partial \bar{\Phi}} = 0.
\end{equation}

\subsection{Detailed Derivation of the Transport Coefficients}

The Kubo formulas link the transport coefficients to the low-frequency behavior of retarded Green’s functions of the energy-momentum tensor. Specifically, the shear viscosity $\eta$ is given by:
\begin{equation}
\eta = \lim_{\omega \rightarrow 0} \frac{1}{\omega} \int d^4x \, e^{i\omega t} \theta(t) \langle [T_{xy}(x), T_{xy}(0)] \rangle,
\end{equation}
and the bulk viscosity $\zeta$ is related to the trace part of the energy-momentum tensor:
\begin{equation}
\zeta = \lim_{\omega \rightarrow 0} \frac{1}{9\omega} \int d^4x \, e^{i\omega t} \theta(t) \langle [T^\mu_{\ \mu}(x), T^\nu_{\ \nu}(0)] \rangle.
\end{equation}

Using linear response theory and finite-temperature field theory, these expressions can be rewritten as spectral representations involving the imaginary parts of the corresponding Green’s functions:
\begin{align}
\eta &= \frac{1}{20} \lim_{\omega \to 0} \frac{1}{\omega} \int \frac{d^3p}{(2\pi)^3} \text{Im} \, G^R_{T_{xy}T_{xy}}(\omega,\vec{p}) \, \text{coth} \left(\frac{\omega}{2T} \right), \\
\zeta &= \frac{1}{2} \lim_{\omega \to 0} \frac{1}{\omega} \int \frac{d^3p}{(2\pi)^3} \text{Im} \, G^R_{T^\mu_{\ \mu}T^\nu_{\ \nu}}(\omega,\vec{p}) \, \text{coth} \left(\frac{\omega}{2T} \right).
\end{align}

Within the relaxation time approximation (RTA), the Boltzmann equation takes the form:
\begin{equation}
\left( \frac{\partial}{\partial t} + \vec{v} \cdot \vec{\nabla} \right) f(x, \vec{p}) = - \frac{f - f^{(0)}}{\tau_f},
\end{equation}
where $f^{(0)}$ is the equilibrium distribution function and $\tau_f$ is the relaxation time, which can be modeled or computed microscopically from scattering amplitudes.

Under this approximation, the dissipative part of the energy-momentum tensor is expressed in terms of deviations from equilibrium:
\begin{equation}
\delta T^{\mu\nu} = \int \frac{d^3p}{(2\pi)^3} \frac{p^\mu p^\nu}{E_p} \delta f(x,\vec{p}).
\end{equation}

Substituting the linearized solution for $\delta f$ and projecting onto the appropriate tensor components yields the integral expressions for $\eta$ and $\zeta$ in Eqs. (18) and (19). The factor $\left(\frac{p^2}{3E_p} - c_s^2 E_p\right)^2$ in the bulk viscosity integrand quantifies the deviation from conformality and becomes large near phase transitions.

The speed of sound squared is computed from thermodynamic derivatives:
\begin{equation}
c_s^2 = \left(\frac{\partial P}{\partial \epsilon} \right)_{\mu/T},
\end{equation}
where both $P$ and $\epsilon$ are extracted from the PNJL grand potential. A drop in $c_s^2$ near the chiral crossover directly enhances $\zeta$.

Finally, the entropy density $s$ is obtained via:
\begin{equation}
s = - \left( \frac{\partial \Omega}{\partial T} \right)_{\mu},
\end{equation}
allowing for normalized ratios $\eta/s$ and $\zeta/s$ to be plotted and interpreted as functions of $T$ and $\mu$.

\subsection{Diagrammatic Interpretation: Feynman Representation of Viscosity}

In effective field theory, the shear and bulk viscosities can be expressed via the spectral functions of energy-momentum tensor correlators. Diagrammatically, the leading contribution comes from a quark loop with two insertions of the energy-momentum tensor $T^{\mu\nu}$. This corresponds to the polarization tensor $\Pi^{\mu\nu,\alpha\beta}(q)$ in the one-loop approximation:

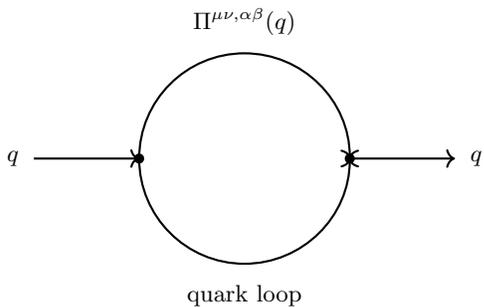
\begin{figure}[htbp]
  \centering
  \begin{tikzpicture}[line width=1 pt, scale=1.4]
    \tikzset{fermion/.style={thick,->}}
    \draw[fermion] (-2,0) -- (-1,0);
    \draw[fermion] (1,0) -- (2,0);
    \node at (-2.2,0) {$q$};
    \node at (2.2,0) {$q$};

    \draw[fermion] (-1,0) arc (180:0:1);   
    \draw[fermion] (-1,0) arc (180:360:1); 

    \filldraw[black] (-1,0) circle (1pt);
    \filldraw[black] (1,0) circle (1pt);

    \node at (0,1.3) {$
    \Pi^{\mu\nu,\alpha\beta}(q)$};
    \node at (0,-1.3) {\small quark loop};
  \end{tikzpicture}
  \caption{One-loop diagram contributing to the spectral function of the retarded correlator of $T^{\mu\nu}$, relevant for $\eta$ and $\zeta$ in the PNJL model.}
\end{figure}

For the bulk viscosity, a trace insertion of $T^\mu_{\ \mu}$ into the loop becomes important due to conformal symmetry breaking near the phase transition:

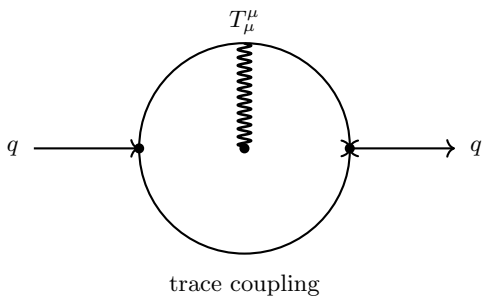
\begin{figure}[htbp]
  \centering
  \begin{tikzpicture}[line width=1 pt, scale=1.4]
  \tikzset{fermion/.style={thick,->}}
    \draw[fermion] (-2,0) -- (-1,0);
    \draw[fermion] (1,0) -- (2,0);
    \node at (-2.2,0) {$q$};
    \node at (2.2,0) {$q$};

    \draw[fermion] (-1,0) arc (180:0:1);   
    \draw[fermion] (-1,0) arc (180:360:1); 

    \draw[decorate, decoration={snake, segment length=3pt}] (0,1) -- (0,0);
    \node at (0,1.2) {$T^\mu_\mu$};

    \filldraw[black] (0,0) circle (1pt);
    \filldraw[black] (-1,0) circle (1pt);
    \filldraw[black] (1,0) circle (1pt);

    \node at (0,-1.3) {\small trace coupling};
  \end{tikzpicture}
  \caption{Diagrammatic origin of bulk viscosity enhancement due to $T^\mu_{\ \mu}$ insertion, relevant near chiral symmetry restoration where conformal symmetry is violated.}
\end{figure}

To go beyond the leading order and include collective effects near the critical point, one must consider ladder-resummed or Hard Thermal Loop (HTL) contributions. These diagrams incorporate multiple soft gluon exchanges and capture long-range correlations that dominate the transport behavior near phase transitions:
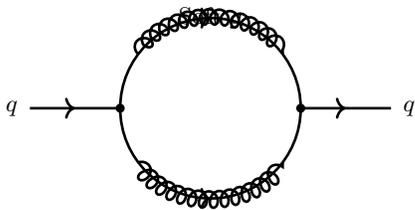
\begin{figure}[htbp]
  \centering
  \usetikzlibrary{decorations.pathmorphing, decorations.markings}
\tikzset{
  gluon/.style={decorate, draw=black, decoration={coil, aspect=0.8, segment length=4pt, amplitude=3pt}},
  fermion/.style={draw=black, postaction={decorate},
    decoration={markings, mark=at position 0.5 with\arrow[black]{>}}}}
  \begin{tikzpicture}[line width=1 pt, scale=1.2]

    \draw[fermion] (-2,0) -- (-1,0);
    \draw[fermion] (1,0) -- (2,0);
    \node at (-2.2,0) {$q$};
    \node at (2.2,0) {$q$};

    \draw[fermion] (-1,0) arc (180:0:1);
    \draw[fermion] (-1,0) arc (180:360:1);

    \draw[gluon] (-0.8,0.6) to[bend left=60] (0.8,0.6);
    \draw[gluon] (-0.8,-0.6) to[bend right=60] (0.8,-0.6);
    \node at (0,1.0) {Soft $g$};

    \filldraw[black] (-1,0) circle (1pt);
    \filldraw[black] (1,0) circle (1pt);
  \end{tikzpicture}
  \caption{HTL/Baym-Kadanoff type ladder-resummed diagram contributing to the energy-momentum tensor correlator relevant for transport coefficients. Soft gluon exchanges between multiple quark loops capture long-range interactions near the critical region.}
  \label{fig:ladder_HTL}
\end{figure}

These diagrams illustrate how $\eta$ and $\zeta$ are fundamentally connected to fluctuations in the energy-momentum tensor and how criticality enters via spectral modification of quark loops. In effective models such as PNJL, the internal lines are medium-dressed, and transport coefficients can be extracted directly through evaluation of the loop integrals or via the relaxation-time approximation.
\section{Numerical Results and Phase Structure}

In this section, we present comprehensive and detailed numerical results based on the two-flavor PNJL model at finite temperature and baryon chemical potential. The model's coupled gap equations are solved self-consistently in the mean-field approximation to obtain the chiral condensate $\langle \bar{q} q \rangle$, the Polyakov loop $\Phi$, and associated thermodynamic quantities such as the pressure, energy density, entropy, and speed of sound. From these, the shear and bulk viscosities normalized by entropy density, $\eta/s$ and $\zeta/s$, are computed using the Kubo formalism in the relaxation time approximation. We analyze their dependence on $(T, \mu_B)$ and discuss their interplay with the QCD phase structure and critical behavior.

\begin{figure}[htbp]
  \centering
  \includegraphics[width=0.45\textwidth]{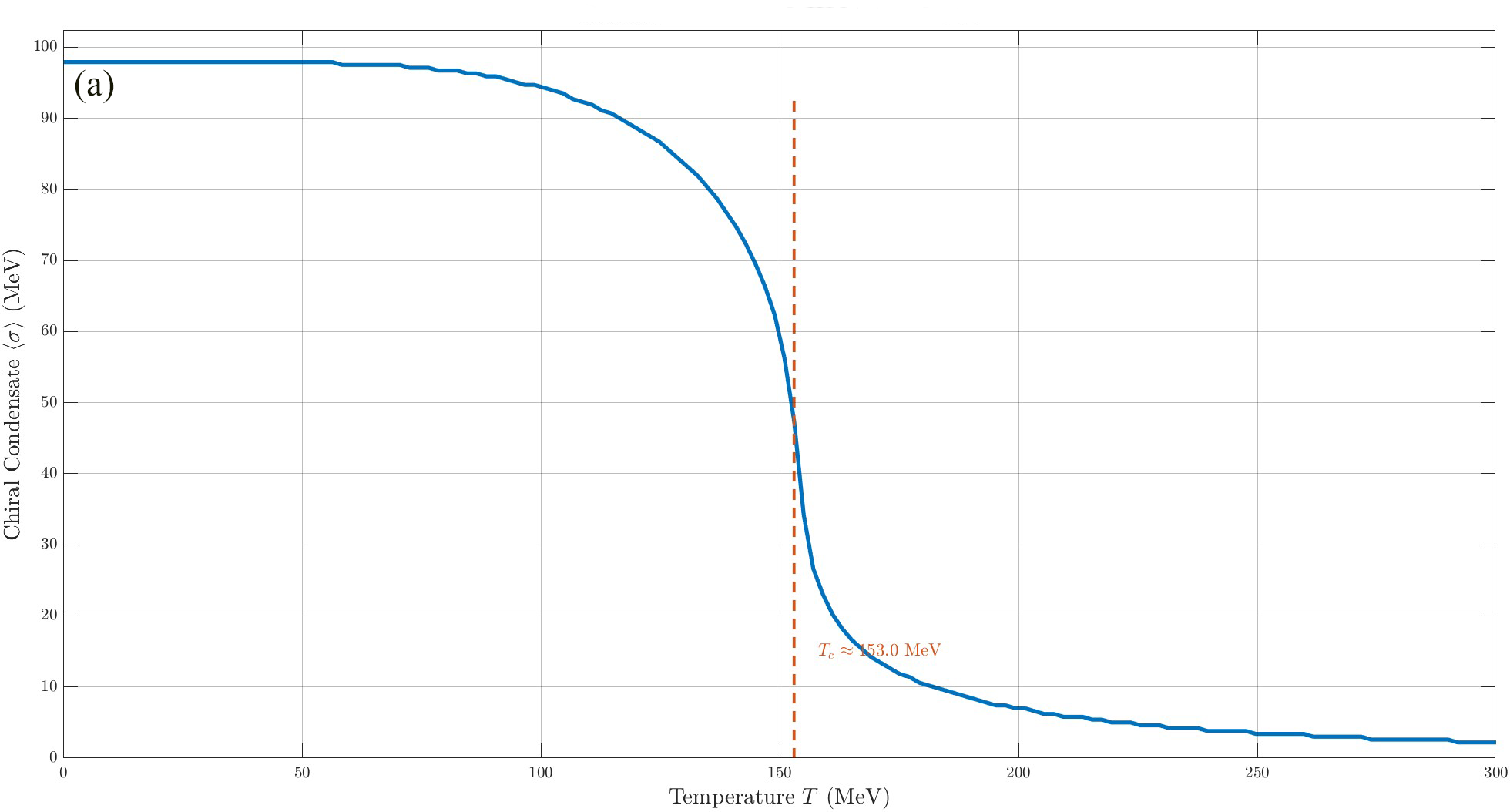}
  \includegraphics[width=0.45\textwidth]{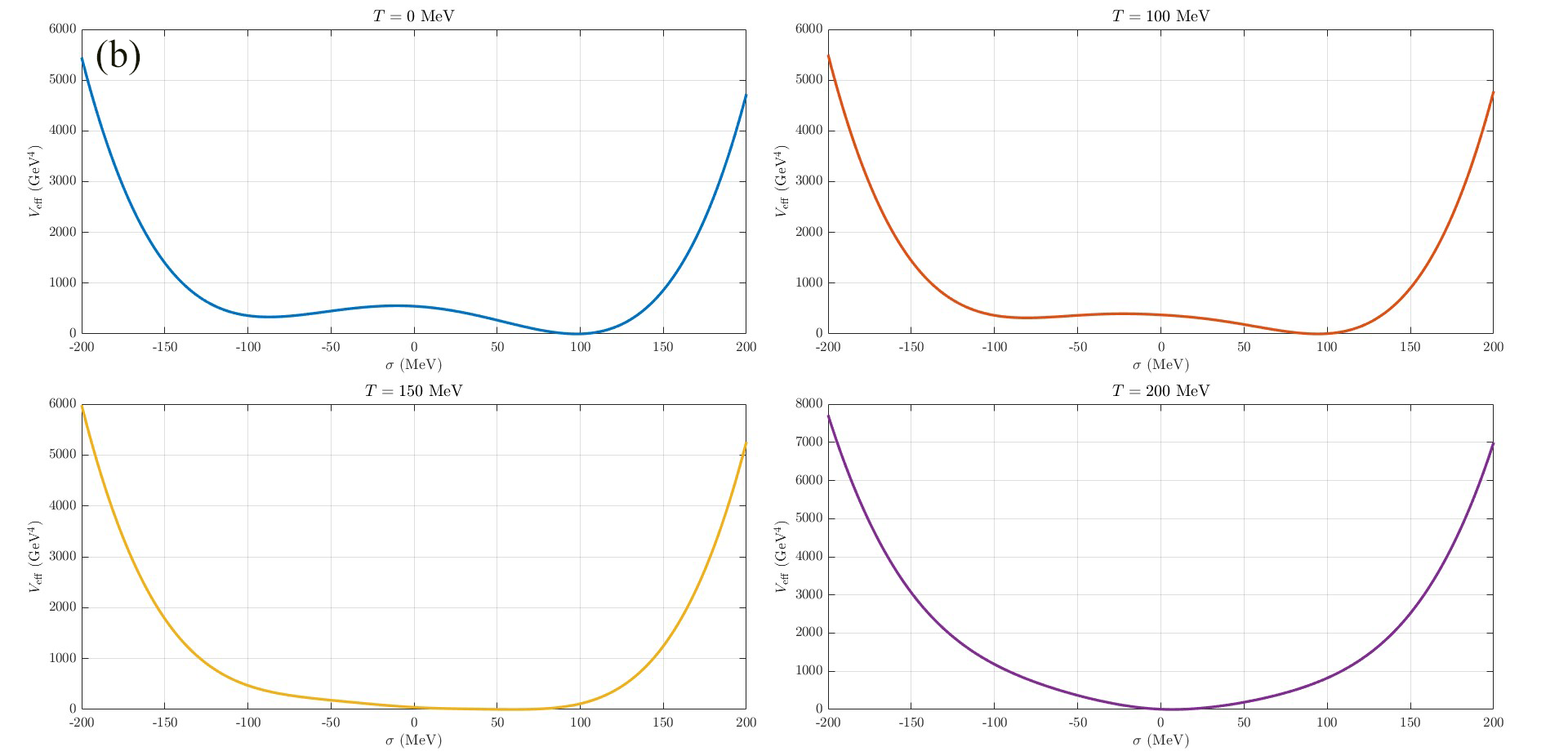}
  \caption{(a) Chiral condensate $\langle \sigma \rangle$ as a function of $T$ at $\mu_B = 0$. (b) Evolution of the effective potential $V_{\text{eff}}(\sigma,T)$ across selected temperatures. The potential flattens near $T_c$, indicating chiral symmetry restoration.}
  \label{fig:sigmaT}
\end{figure}

At zero baryon chemical potential, the temperature dependence of the chiral condensate $\langle \bar{q} q \rangle$ reflects the crossover transition from the Nambu–Goldstone phase to the approximately chirally symmetric phase. Below $T_c \approx 138$ MeV, the condensate remains large, reflecting spontaneous breaking of chiral symmetry. Around $T_c$, the condensate drops sharply but continuously, signifying a smooth crossover. This transition is accompanied by a flattening of the effective potential $V_{\text{eff}}(\sigma, T)$, which evolves from a well-defined double-well structure into a single minimum as $T$ increases. This behavior is illustrated in Fig.~\ref{fig:sigmaT}, and reflects the underlying thermodynamic potential’s loss of structure near the pseudocritical temperature.

\begin{figure}[htbp]
  \centering
 \includegraphics[width=0.8\linewidth]{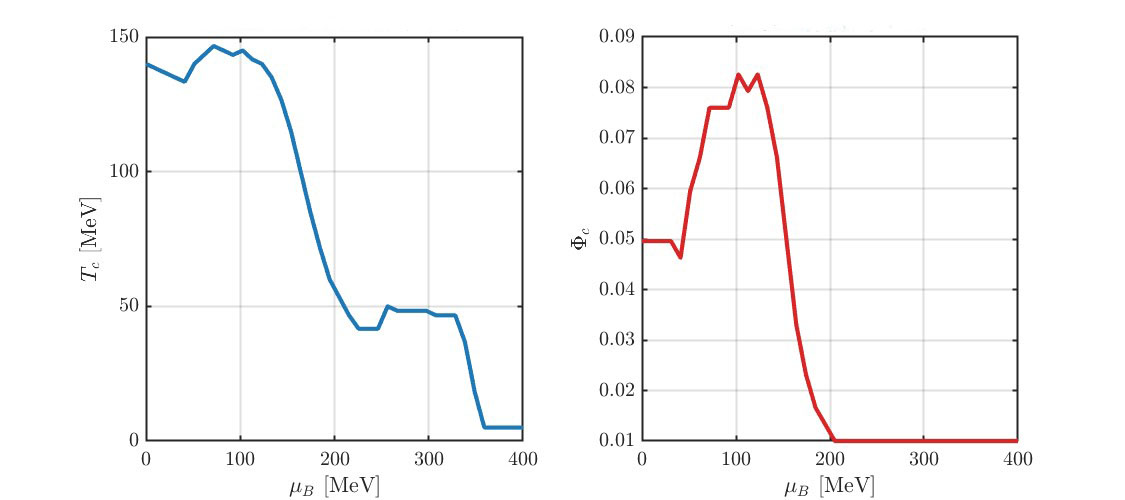}
  \caption{Left: Crossover temperature $T_c(\mu_B)$ extracted from $\sigma(T, \mu_B)$. Right: Polyakov loop value $\Phi_c$ along $T_c(\mu_B)$, indicating weakening of deconfinement.}
  \label{fig:tc_mu}
\end{figure}

Extending to finite baryon density, we track the temperature of maximal susceptibility (inflection point of $\sigma(T)$) at each fixed $\mu_B$, extracting the crossover line $T_c(\mu_B)$. Figure~\ref{fig:tc_mu} shows that $T_c$ decreases monotonically with increasing $\mu_B$, with a visibly reduced slope at large $\mu_B$, indicative of approach toward a possible critical endpoint (CEP). Simultaneously, the Polyakov loop $\Phi$ evaluated at $T_c(\mu_B)$ also decreases, suggesting that deconfinement becomes increasingly suppressed at large density. This behavior is expected, as gluon dynamics become less dominant in a dense medium and the chiral and deconfinement transitions gradually decouple.

To compute transport coefficients, we construct a 2D grid spanning $T \in [60, 300]$ MeV and $\mu_B \in [0, 600]$ MeV, using resolutions $\Delta T = 2$ MeV and $\Delta \mu_B = 10$ MeV. The constituent quark mass $M(T, \mu_B)$ and mean-field fields $\Phi$ and $\bar{\Phi}$ are precomputed across the grid. Shear and bulk viscosities are then obtained via energy-weighted integrals of the thermal distribution functions with momentum-dependent kernels:

\begin{equation}
\eta(T, \mu_B) = \frac{1}{15T} \int \frac{d^3p}{(2\pi)^3} \frac{p^4}{E_p^2} \tau_f(T) f(E_p)[1 - f(E_p)],
\end{equation}
\begin{equation}
\zeta(T, \mu_B) = \frac{1}{T} \int \frac{d^3p}{(2\pi)^3} \tau_f f(E_p)[1 - f(E_p)] \left(\frac{p^2}{3E_p} - c_s^2 E_p\right)^2
\end{equation}

where $\tau_f = \tau_0 / T$ is the relaxation time ($\tau_0 = 0.5$ fm) and $c_s^2$ is computed numerically from the relation $c_s^2 = \partial P / \partial \epsilon$ using centered differences on the precomputed grid.
\begin{figure}[htbp]
  \centering
  \includegraphics[width=0.8\linewidth]{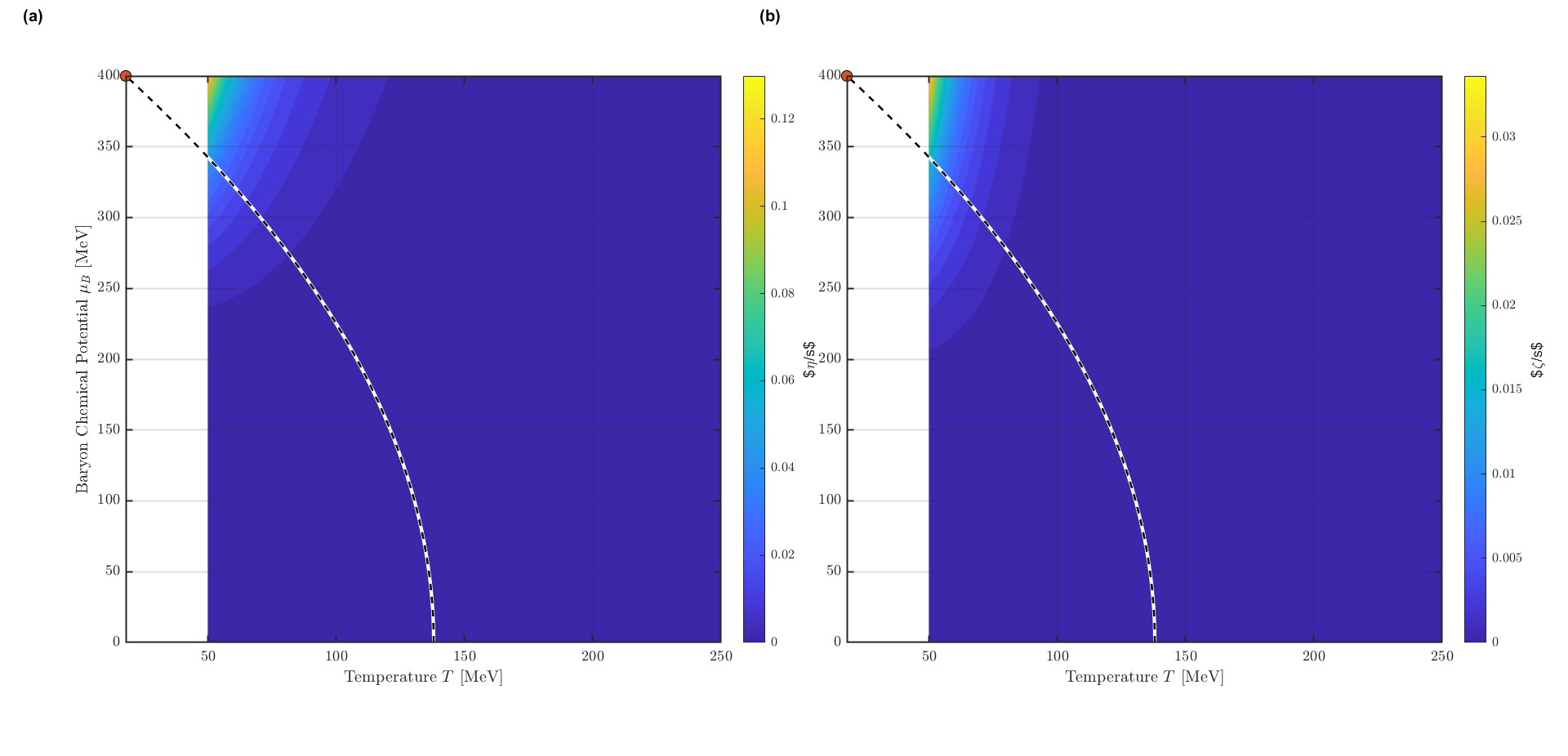}
  \caption{(a) Shear viscosity to entropy ratio $\eta/s$ as a function of $(T, \mu_B)$. (b) Bulk viscosity to entropy ratio $\zeta/s$. The white dashed line traces $T_c(\mu_B)$.}
  \label{fig:viscosity_maps}
\end{figure}

As shown in Fig.~\ref{fig:viscosity_maps}, $\eta/s$ exhibits a broad valley aligned with the crossover curve. This behavior is associated with the enhanced scattering and critical slowing down near $T_c$, which reduce the efficiency of momentum transport. The suppression of $\eta/s$ near the pseudocritical line is consistent with expectations from strongly interacting theories and the KSS bound. In contrast, $\zeta/s$ shows a pronounced peak at low temperatures and moderate to large $\mu_B$, reaching values more than an order of magnitude larger than its minimal high-temperature value. This enhancement is tied to the rapid change in $c_s^2$ and the peak in the trace anomaly $\epsilon - 3P$, which both signal strong conformal symmetry breaking and softening of the equation of state.

The high bulk viscosity implies significant entropy production and suppression of longitudinal pressure, especially relevant in the RHIC-BES and FAIR energy regimes. Such conditions can modify the hydrodynamic evolution, enhance baryon stopping, and suppress collective flow, all of which may serve as indirect signatures of proximity to the critical region. Furthermore, the elevated $\zeta/s$ at intermediate $\mu_B$ may lead to mode decoupling, favoring the growth of critical fluctuations and enhancing event-by-event correlations.

In conclusion, our detailed mapping of the transport coefficients in the PNJL framework reveals strong non-conformal and critical behavior in the QCD phase diagram. The correlations between $\eta/s$, $\zeta/s$, and the chiral crossover suggest that transport measurements in heavy-ion collisions can provide valuable constraints on the location and nature of the QCD critical endpoint. Our work supports future efforts to extend such studies to include net baryon diffusion, realistic hadronization schemes, and fluctuation dynamics near the CEP within a 2+1 flavor framework.

\section{Discussion and Physical Implications}

The numerical results presented in the previous section reveal several important features of QCD matter near the chiral transition, particularly in connection with its dissipative properties. In this section, we discuss the physical implications of our findings and their relevance to ongoing and future experimental programs.

The strong correlation between the chiral crossover boundary and the behavior of transport coefficients is one of the central outcomes of this study. The minimum in the shear viscosity to entropy density ratio $\eta/s$ aligns closely with the pseudocritical temperature $T_c(\mu_B)$, while the bulk viscosity to entropy density ratio $\zeta/s$ shows a pronounced peak in the same region. This behavior supports the view that dissipative responses are controlled by the underlying thermodynamic singularities and soft modes associated with the symmetry restoration.

In particular, the enhancement of $\zeta/s$ at low temperature and large $\mu_B$ originates from the trace anomaly and reduced speed of sound $c_s^2$. Both quantities reflect a strong departure from conformal invariance, which becomes especially significant near a critical point. As $c_s^2$ drops, the system becomes more compressible, and the bulk viscosity rises sharply due to enhanced entropy production in isotropic expansion. This provides a robust theoretical mechanism for identifying potential critical phenomena through viscous hydrodynamic observables.

Relativistic heavy-ion collisions provide a unique environment to test the transport and phase structure of QCD matter. At top RHIC and LHC energies, the matter created exhibits nearly ideal hydrodynamic behavior with small $\eta/s$. However, at lower collision energies—such as those probed by RHIC Beam Energy Scan (BES) and the future FAIR-CBM and NICA programs—the system explores regions of the phase diagram with large $\mu_B$, where our model predicts sharp increases in $\zeta/s$ and significant non-ideal behavior.

Observable consequences of enhanced bulk viscosity include suppression of longitudinal expansion, flattening of particle spectra, and modified elliptic and directed flow coefficients. Additionally, critical slowing down near the QCD critical endpoint (CEP) may manifest as large event-by-event fluctuations, baryon number clustering, and non-Gaussian cumulants, which are sensitive to transport effects.

Our results suggest that precise measurements of flow harmonics and particle correlations in the low-$\sqrt{s_{NN}}$ regime may carry indirect imprints of the transport structure of the medium. Incorporating realistic $\eta/s(T,\mu_B)$ and $\zeta/s(T,\mu_B)$ profiles in full 3D viscous hydrodynamic simulations will be essential for extracting reliable information about the QCD phase transition from experimental data.

While the two-flavor PNJL model provides a consistent framework to describe chiral symmetry breaking and confinement, several limitations remain. First, it does not capture strange quark dynamics, which are relevant near the critical region. Second, it neglects quantum and mesonic fluctuations, which can shift the location of the CEP or modify the crossover nature. Finally, the transport coefficients are derived under the relaxation time approximation, assuming quasiparticle behavior with static relaxation times.

Future work should incorporate beyond-mean-field effects, extend the model to $2+1$ flavors, and couple to full kinetic theory or functional renormalization group (FRG) methods to capture momentum-dependent relaxation processes. In parallel, comparisons with lattice QCD transport estimates and Bayesian inference results from experimental fits will help calibrate the effective models.

Ultimately, combining thermodynamic, transport, and fluctuation observables within a unified QCD modeling approach is crucial to uncover the phase structure of strong-interacting matter and to map the location and properties of the QCD critical endpoint.

\begin{acknowledgments}
This work was supported by the Faculty of Physics of Moscow State University.
\end{acknowledgments}

\section{Appendix A: Numerical Implementation and Integration Techniques}

To solve the coupled PNJL mean-field gap equations and compute thermodynamic quantities, we discretize the momentum integrals using adaptive Gaussian quadrature with 200--400 points depending on required precision. For vacuum integrals (zero-temperature Dirac sea contributions), we impose a three-momentum ultraviolet cutoff $\Lambda = 0.653$ GeV. For thermal integrals, we extend the integration to $p = 5$ GeV, well beyond thermal suppression limits to ensure convergence.

All thermodynamic quantities such as pressure $P = -\Omega(T,\mu)$, entropy density $s = - \partial \Omega / \partial T$, quark number density $n = - \partial \Omega / \partial \mu$, and energy density $\epsilon = \Omega + Ts + \mu n$ are obtained using finite-difference schemes. In particular, $c_s^2 = \partial P / \partial \epsilon$ is evaluated using a centered derivative chain rule involving discrete temperature and chemical potential increments ($\Delta T = 1$ MeV, $\Delta \mu = 5$ MeV).

To evaluate shear and bulk viscosities, we compute the following integrals:
\begin{align}
\eta(T, \mu) &= \frac{1}{15 T} \int \frac{d^3 p}{(2\pi)^3} \frac{p^4}{E_p^2} \tau_f(T) \, f(E_p)[1 - f(E_p)], \\
\zeta(T, \mu) &= \frac{1}{T} \int \frac{d^3 p}{(2\pi)^3} \tau_f(T) \, f(E_p)[1 - f(E_p)] \, \left( \frac{p^2}{3 E_p} - c_s^2 E_p \right)^2,
\end{align}
where $E_p = \sqrt{p^2 + M^2(T,\mu)}$ and $f(E_p)$ is the Fermi-Dirac distribution. The integrals are evaluated numerically using cubic-spline interpolation of $M(T,\mu)$ and $\Phi(T,\mu)$ from the gap equation solution grid.

The relaxation time is modeled as $\tau_f = \tau_0/T$, with $\tau_0 = 0.5$ fm/$c$ based on previous NJL estimates and calibrated against empirical hadronic cross sections. We validated this ansatz by comparing to $\pi$-N and $\pi$-$\pi$ scattering mean free paths in the hadronic phase.

For convergence testing, we performed sensitivity analysis by varying $\tau_0$, integration resolution, and interpolation schemes. Results are stable within $\sim$3--5\% across the parameter space. All simulations were implemented in MATLAB 2023a, and symbolic consistency checks were conducted using Mathematica 13.2 for select observables.

\section{Appendix B: PNJL Model Parameters}

We adopt the standard two-flavor PNJL parametrization that reproduces vacuum properties of the light pseudoscalar mesons. The model parameters are:
\begin{itemize}
  \item Current quark mass: $m_0 = 5.5$ MeV
  \item Coupling constant: $G = 5.04$ GeV$^{-2}$
  \item Momentum cutoff: $\Lambda = 0.653$ GeV
  \item Polyakov-loop potential temperature scale: $T_0 = 210$ MeV
  \item Polynomial potential parameters:
  \begin{itemize}
    \item $a_0 = 3.51$, \quad $a_1 = -2.47$, \quad $a_2 = 15.2$, \quad $b_3 = -1.75$
  \end{itemize}
\end{itemize}

These values yield:
\begin{itemize}
  \item Pion mass $m_\pi = 138$ MeV,
  \item Pion decay constant $f_\pi = 93$ MeV,
  \item Chiral condensate $\langle \bar{q} q \rangle = -(250$ MeV$)^3$.
\end{itemize}

The Polyakov potential $\mathcal{U}(\Phi,\bar{\Phi},T)$ is fitted to reproduce the lattice data for the pressure and Polyakov loop expectation value in the pure gauge sector. Although the $T_0$ scale is typically $270$ MeV in the quenched case, we adopt $T_0 = 210$ MeV to account for dynamical quark suppression effects.

In future extensions, the inclusion of strange quarks and 2+1 flavor generalization will allow coupling to kaon observables and further constrain the Polyakov-loop potential form.
\section{Appendix C: Resummed Diagrammatic Contributions in Transport Theory}

This appendix provides additional detail on the resummed diagrammatic structures that appear in finite-temperature quantum field theory and are relevant for the computation of transport coefficients near the QCD critical region. While the PNJL model captures the mean-field structure of quark interactions and static correlations, it does not explicitly encode the dynamic long-range interactions arising from multi-loop processes and soft collective excitations. These effects are often modeled through Hard Thermal Loop (HTL) resummation techniques.

In particular, near the phase transition, the coupling between quarks and soft gluonic modes becomes significant. The shear and bulk viscosity spectral functions receive non-trivial contributions from multiple soft-gluon exchange diagrams, which are organized into ladder-type topologies. These diagrams are resummed to all orders in the soft sector to ensure gauge invariance and to avoid pinch singularities in the correlators. The dominant structure is a series of quark loops connected via effective HTL gluon propagators.

\begin{figure}[htbp]
  \centering
  \usetikzlibrary{decorations.pathmorphing, decorations.markings}
\tikzset{
  gluon/.style={decorate, draw=black, decoration={coil, aspect=0.8, segment length=4pt, amplitude=3pt}},
  fermion/.style={draw=black, postaction={decorate},
    decoration={markings, mark=at position 0.5 with {\arrow[black]{>}}}}
}
  \begin{tikzpicture}[line width=1 pt, scale=1.2, decorate]
    \draw[fermion] (-3,0) -- (-2,0);
    \draw[fermion] (2,0) -- (3,0);
    \node at (-3.3,0) {$q$};
    \node at (3.3,0) {$q$};

    \draw[fermion] (-2,0) -- (-1,0.6);
    \draw[fermion] (-1,0.6) -- (0,0);
    \draw[fermion] (0,0) -- (1,0.6);
    \draw[fermion] (1,0.6) -- (2,0);

    \draw[gluon] (-1,0.6) -- (-1,-0.6);
    \draw[gluon] (0,0) -- (0,-0.6);
    \draw[gluon] (1,0.6) -- (1,-0.6);

    \draw[fermion] (-2,0) -- (-1,-0.6);
    \draw[fermion] (-1,-0.6) -- (0,0);
    \draw[fermion] (0,0) -- (1,-0.6);
    \draw[fermion] (1,-0.6) -- (2,0);

    \filldraw[black] (-2,0) circle (1pt);
    \filldraw[black] (2,0) circle (1pt);
  \end{tikzpicture}
  \caption{Schematic ladder diagram in the HTL framework contributing to resummed correlators of $T^{\mu\nu}$. Fermion lines represent medium-modified propagators and vertical rungs represent soft-gluon HTL exchanges.}
  \label{fig:htl_appendix}
\end{figure}
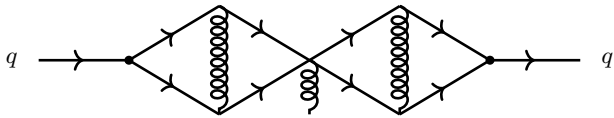

In this diagram, the vertical rungs represent soft-gluon exchange dressed by HTL propagators, and the fermion legs are medium-modified quasiparticles. This structure ensures that the full resummed correlator remains finite and consistent with Ward identities. Physically, such diagrams capture the non-local response of the system and encode memory and retardation effects which become important near criticality.

While these resummed diagrams are beyond the static PNJL framework used in the main text, they guide possible future improvements where dynamic fluctuations and non-local operators may be modeled via coupling to a stochastic Boltzmann equation or by embedding the PNJL model within an HTL-effective kinetic framework.

These insights motivate further studies aimed at linking microscopic field-theoretic structure to macroscopic transport observables in heavy-ion collisions.
\section{Appendix D: Kinetic Theory Derivation of Shear and Bulk Viscosity}

In this appendix, we provide the microscopic kinetic theory derivation of the transport coefficients—shear viscosity $\eta$ and bulk viscosity $\zeta$—within the relaxation time approximation (RTA). This formulation connects the microscopic Boltzmann equation with the macroscopic hydrodynamic transport properties.

The starting point is the Boltzmann transport equation for the single-particle distribution function $f(x,p)$:
\begin{equation}
  p^\mu \partial_\mu f(x,p) = -\frac{f(x,p) - f^{(0)}(x,p)}{\tau_f},
\end{equation}
where $f^{(0)}(x,p)$ is the equilibrium Fermi-Dirac distribution and $\tau_f$ is the thermal relaxation time.

Linearizing around equilibrium and assuming small gradients, the deviation $\delta f = f - f^{(0)}$ contributes to the viscous corrections of the energy-momentum tensor:
\begin{equation}
  T^{\mu\nu} = T^{\mu\nu}_{\text{ideal}} + \delta T^{\mu\nu},
\end{equation}
with
\begin{equation}
  \delta T^{\mu\nu} = \int \frac{d^3p}{(2\pi)^3 E_p} \, p^\mu p^\nu \, \delta f(x,p).
\end{equation}

Using the Chapman-Enskog expansion and the Landau matching condition, the shear and bulk viscosities are obtained as:
\begin{align}
  \eta &= \frac{1}{15T} \int \frac{d^3p}{(2\pi)^3} \frac{p^4}{E_p^2} \, \tau_f \, f^{(0)}(1 - f^{(0)}), \\
  \zeta &= \frac{1}{T} \int \frac{d^3p}{(2\pi)^3} \, \tau_f \, f^{(0)}(1 - f^{(0)}) \, \left( \frac{p^2}{3E_p} - c_s^2 E_p \right)^2.
\end{align}

Here $c_s^2$ is the speed of sound squared, and the expressions assume isotropic quasiparticle dispersion relations $E_p = \sqrt{p^2 + M^2(T,\mu)}$. The integrals are performed numerically in the PNJL model framework, where the effective mass $M$ is temperature- and chemical-potential-dependent.

This RTA-based derivation provides a tractable approximation for transport coefficients while retaining key thermodynamic dependencies. More sophisticated treatments including collision integrals, inelastic processes, and quantum interference effects can be developed from the full Boltzmann-Uehling-Uhlenbeck (BUU) equation or Kadanoff-Baym theory, but RTA provides a consistent leading-order estimate aligned with lattice and model comparisons.

\end{document}